\documentclass[12pt]{article}
\usepackage{amsmath,amssymb,cite}
\numberwithin{equation}{section}

\begin{document}

\title{\begin{flushright}
\small SU-4252-836\\IMSc/2006/08/19\\IISc/CHEP/6/06
\end{flushright}
\vspace{0.25cm} Statistics and UV-IR Mixing with Twisted Poincar\'{e}
Invariance} 
\author{A. P. Balachandran$^a$\footnote{bal@phy.syr.edu} ,T. R.
Govindarajan$^b$\footnote{trg@imsc.res.in},
G. Mangano$^c$\footnote{mangano@na.infn.it},\\
A. Pinzul$^a$\footnote{apinzul@phy.syr.edu} ,
B. A. Qureshi$^a$\footnote{bqureshi@phy.syr.edu}
and S. Vaidya$^d$\footnote{vaidya@cts.iisc.ernet.in}\\ \\
$^a$\begin{small}Department of Physics, Syracuse University, Syracuse NY,
13244-1130, USA. \end{small} \\
$^b$\begin{small}The Institute of Mathematical Sciences,
C.I.T. Campus, Taramani, Chennai 600 113, India.\end{small}\\
$^c$\begin{small}INFN, Sezione di Napoli and Dipartimento di Scienze
  Fisiche,\end{small}\\ 
\begin{small}Universit\`{a} di Napoli {\it Federico II}, Via Cintia,
I-80126 Napoli, Italy.\end{small}\\
$^d$\begin{small}Centre for High Energy Physics, Indian
Institute of Science,
Bangalore, 560012, India.
\end{small}}
\date{\empty}

\maketitle

\begin{abstract}
We elaborate on the role of quantum statistics in twisted Poincar\'e
invariant theories. It is shown that, in order to have twisted
Poincar\'e group as the symmetry of a quantum theory, statistics must
be twisted. It is also confirmed that the removal of UV-IR mixing (in
the absence of gauge fields) in such theories is a natural
consequence.
\end{abstract}

\section{Introduction}
Following the application of Drinfel'd's twist for the Poincar\'e
group on the noncommutative Groenewold-Moyal (GM) plane
\cite{Chaichian:2004za,Aschieri:2005yw}, much interest has been
generated in the study of its physical consequences. One such
consequence pointed out in \cite{oeckl1,bmpv} is that the usual
statistics is not compatible with the twisted action of the Poincar\'e
group. This is in agreement with what is already known in quantum
group theory. Among the consequences of this result is the removal of
UV-IR mixing \cite{bpq} in the $S$-matrix in the absence of gauge
fields.

Recently there have been claims that this twisting of statistics is
unnecessary or even wrong, and that the removal of UV-IR mixing is the
result of a wrong choice of interaction. In this note we explain our
point of view more clearly, demonstrating that if one wants to retain
the twisted Poincar\'e symmetry in a quantum theory, then one is forced
to implement twisted statistics. Secondly, the form of the interaction
is dictated by quantum symmetry as well.

The paper is organized as follows. After briefly reviewing the
Drinfel'd twist for Poincar\'e group in the section 2, we elaborate on
its implications for quantum statistics in section 3. Section 4
discusses the choice of the correct twisted Lorentz-invariant
interaction Hamiltonian. In section 5, we show by an explicit
calculation that the correlation functions and hence the $S$-matrix of
the noncommutative quantum field theory (NCQFT) with usual statistics
are not invariant under the twisted symmetry, while the same are
manifestly so for the theory with twisted statistics. Section 6
discusses some issues related to the functional integral for theories
with twisted Poincar\'e symmetry.  Section 7 describes the notion of
locality in the twisted statistics approach and Section 8 addresses
some general issues regarding the tensor products of fields.

\section{The Twist}

The action of a symmetry group on the tensor product of representation
spaces carrying the same representation $\rho$ is given by a
coproduct $\Delta$:
\begin{equation}
g \triangleright\,(\phi\otimes\chi) = (\rho\otimes\rho)
\Delta(g)\, (\phi\otimes\chi).
\end{equation}
If the representation space happens to be an algebra as well, we further
have the compatibility condition
\begin{equation}
m\big((\rho\otimes\rho) \Delta(g) (\phi\otimes\chi)\big) = \rho(g)\,m
(\phi\otimes\chi)
\end{equation}
where $m$ is the multiplication map.

The GM plane is the algebra ${\cal A}_\theta$ of functions $f
\in\mathbb R^n$ with the product defined by
\begin{equation}
f\,\ast\,g\,=\,m_\theta (f\otimes g) =\,m_0 (\mathcal F \,f\otimes
g)
\label{stardef}
\end{equation}
where $m_0$ is the usual point-wise multiplication,
\begin{equation}
\mathcal F\,=\,e^{-\frac{i}{2}\theta^{\mu\nu}P_\mu\otimes P_\nu}, \quad
P_\mu = -i \partial_\mu \, ,
\end{equation}
is called the {\it twist} element, and this rule for multiplication is
often called the {\it star} product. Explicitly (\ref{stardef}) gives
\begin{equation} 
(f \ast g)(x) = \left.\exp \left(\frac{i}{2}\theta^{\mu \nu}
  \frac{\partial}{\partial x^\mu}\frac{\partial}{\partial y^\nu}
  \right)f(x)g(y)\right|_{x=y} \, . 
\end{equation}  

The usual coproduct $\Delta_0$ on the Poincar\'e group,
\begin{equation} 
\Delta_0(\Lambda) = \Lambda \times \Lambda, \quad \Lambda \in
\text{Poincar\'e \,\,group}, 
\end{equation} 
 is not compatible with the star product. But a new coproduct
$\Delta_\theta$ obtained using the twist is compatible, where
\begin{equation}
\Delta_\theta(\Lambda)\,=\,\mathcal F^{-1}\,\Delta_0 (\Lambda) \mathcal F.
\label{twistedrep}
\end{equation}

For details see \cite{Chaichian:2004za,Aschieri:2005yw}. Note that
$\Delta_\theta(a)=\Delta_0(a)$ if $a$ is a translation.

\section{Twisted Statistics}
Twisting the coproduct implies twisting of statistics in quantum
theory, as we will argue in this section. This result holds for an
$n$-particle quantum mechanical system and also for quantum field
theory.

\subsection{Quantum Mechanics}
The wave function of a two-particle system for $\theta^{\mu \nu}=0$ in
position representation is a function of two variables, hence lives in
$\mathcal A_0 \otimes \mathcal A_0$, the tensor product of two copies
of the algebra of functions on commutative ${\mathbb R}^n$, and
transforms according to the usual coproduct $\Delta_0$. Similarly in
noncommutative case, the wavefunction lives in $\mathcal A_\theta
\otimes \mathcal A_\theta$ and transforms according to the twisted
coproduct $\Delta_\theta$.

A general element of the tensor product has no particular
symmetry. Usually we require that the physical wave functions
describing identical particles are either symmetric (bosons) or
antisymmetric (fermions). This requires us to work with either the
symmetrized or antisymmetrized tensor product
\begin{eqnarray} 
\phi \otimes_S \chi &\equiv& \frac{1}{2}\left(\phi \otimes \chi +\chi
\otimes \phi\right) \, ,\label{symm}\\
\phi \otimes_A \chi &\equiv& \frac{1}{2}\left(\phi \otimes \chi -\chi
\otimes \phi \right) \label{asymm}
\end{eqnarray} 
which satisfy
\begin{eqnarray} 
\phi\,\otimes_S\chi = +\chi\,\otimes_S \phi \, , \\
\phi\,\otimes_A\chi = -\chi\,\otimes_A \phi\, . 
\end{eqnarray} 
In a Lorentz-invariant theory, these relations have to hold in all
frames of reference. In other words, performing a Lorentz
transformation on $\phi \otimes \chi$ and then (anti-)symmetrizing has
to be the same as (anti-)symmetrization followed by the Lorentz
transformation.

It is not difficult to show that the twisted coproduct
(\ref{twistedrep}) is not compatible with usual
symmetrization/antisymmetrization (\ref{symm}, \ref{asymm}). To see
this, let us write $\mathcal F^{-1}$ and $\mathcal F$ in the Sweedler
notation (see for e.g. page 5 of \cite{majid}) as
\begin{eqnarray} 
\mathcal F^{-1}\,&=& \sum_\alpha f^{(1)\alpha} \otimes f^{(2)}_\alpha, \quad
\mathcal F\,= \sum_\alpha\tilde{f}^{(1)\alpha} \otimes
\tilde{f}^{(2)}_\alpha, \quad {\rm with} \\ 
\mathcal F^{-1} \mathcal F &=& {\bf 1} \otimes {\bf 1} =
\sum_{\alpha,\beta} f^{(1)\alpha}\tilde{f}^{(1)\beta} \otimes
f^{(2)}_\alpha \tilde{f}^{(2)}_\beta \, .\label{inentity}
\end{eqnarray} 
Under a Lorentz transformation $\Lambda$,
\begin{eqnarray}
\Lambda: \phi\,\otimes\,\chi\ &\longrightarrow & (\rho \otimes \rho)
\Delta_\theta(\Lambda)(\phi\,\otimes\,\chi) \nonumber \\
&=&\sum_{\alpha,\beta} \rho(f^{(1)\alpha} \Lambda
\tilde{f}^{(1)\beta}) \phi \otimes \,\rho(f^{(2)}_\alpha \Lambda 
\tilde{f}^{(2)}_\beta) \chi\,.
\label{Ltrans} 
\end{eqnarray} 
Subsequent symmetrization/antisymmetrization gives us
\begin{equation} 
\sum_{\alpha,\beta} \big(\rho(f^{(1)\alpha} \Lambda \tilde{f}^{(1)\beta})\phi
\,\otimes \,\rho(f^{(2)}_\alpha \Lambda \tilde{f}^{(2)}_\beta) \chi \pm
\rho(f^{(2)}_\alpha \Lambda \tilde{f}^{(2)}_\beta) \chi
\otimes \rho(f^{(1)\alpha} \Lambda \tilde{f}^{(1)\beta})\phi \big)
\label{Lsymm}
\end{equation} 
whereas 
\begin{eqnarray} 
\lefteqn{(\rho \otimes \rho) \Delta_\theta (\Lambda) (\phi \otimes_{S,A}
  \chi) = } \nonumber \\
&&\sum_{\alpha'\beta} \big(\rho(f^{(1)\alpha} \Lambda
\tilde{f}^{(1)\beta})\phi \,\otimes \,\rho(f^{(2)}_\alpha \Lambda
\tilde{f}^{(2)}_\beta) \chi \pm \rho(f^{(1)\alpha} \Lambda
  \tilde{f}^{(1)\beta})\chi \otimes \,\rho(f^{(2)}_\alpha \Lambda
  \tilde{f}^{(2)}_\beta) \phi \big) 
\label{symmL}
\end{eqnarray} 
which is not the same as (\ref{Lsymm}) [See \cite{bmpv} for the same
proof which avoids Sweedler notation.].  The origin of this difference
can be traced to the fact that the coproduct is not cocommutative
except when $\theta^{\mu \nu} =0$.

There is another way to phrase this compatibility (or lack thereof) of
Lorentz transformations and symmetrization. Let $\tau_0$
be the statistics (flip) operator associated with exchange:
\begin{equation}
\tau_0(\phi \otimes \chi) = \chi \otimes \phi\, .
\end{equation}
In usual quantum theory, we have the axiom that $\tau_0$ is
superselected, i.e., all the observables commute with $\tau_0$. What
this means is that no operator in the physical Hilbert space can
change statistics. In particular the quantum operators that implement
Lorentz symmetry must commute with the statistics operator. Also all
the states in a given superselection sector are eigenstates of
$\tau_0$ with the same eigenvalue. Given an element
$\phi\,\otimes\,\chi$ of the tensor product, the physical Hilbert
spaces can be constructed from the elements
\begin{equation} 
\left(\frac{1 \pm \tau_0}{2}\right)\,(\phi\,\otimes\,\chi)\, .
\end{equation} 
As is obvious from eq (\ref{Lsymm},\ref{symmL}),
\begin{equation}
\tau_0\ \Delta_\theta(\Lambda) \neq\ \Delta_\theta(\Lambda) \tau_0
\end{equation}
showing that the usual statistics is not compatible with the
coproduct. But notice that the new statistics operator
\begin{equation}
\tau_\theta\,\equiv \,\mathcal F^{-1}\,\tau_0 \mathcal F, \quad
\tau_\theta^2 = {\bf 1}\otimes {\bf 1} 
\label{newstat}
\end{equation}
does commute with the twisted coproduct. The states constructed according to
\begin{equation} 
\phi \otimes_{S_\theta} \chi \equiv
\left(\frac{1\,+ \tau_\theta}{2}\right)\,(\phi\,\otimes\,\chi), \quad
\phi \otimes_{A_\theta} \chi \equiv \left(\frac{1\,-
  \tau_\theta}{2}\right)\,(\phi\,\otimes\,\chi)
\end{equation} 
form the physical two-particle Hilbert spaces of (generalized) bosons
and fermions and obey twisted statistics.

For plane waves $e_p(x) = e^{-i p \cdot x}$ we get
\begin{eqnarray} 
\left(\frac{{\bf 1} \pm \tau_\theta}{2}\right)(e_p \otimes e_q) &\equiv&
e_p \otimes_{S_\theta,A_\theta} e_q = \pm e^{-i p_\mu
  \theta^{\mu \nu} q_\nu} e_q \otimes_{S_\theta,A_\theta} e_p \, ,\\
(e_p \otimes_{S_\theta,A_\theta} e_q)(x_1,x_2) &=& \pm e^{-i
  \frac{\partial}{\partial x^\mu_1} \theta^{\mu \nu}
  \frac{\partial}{\partial x^\nu_2}} (e_p \otimes_{S_\theta,A_\theta}
  e_q)(x_2,x_1) \, . \label{tplane}
\end{eqnarray} 

Using the anti-symmetry of $\theta^{\mu \nu}$, $\tau_\theta$ may also
be equivalently written as
\begin{equation} 
\tau_\theta = \mathcal F^{-2} \tau_0 \, .
\end{equation} 
This form of $\tau_\theta$ allows to make contact with quantum group
theory, and identifies $\mathcal F^{-2}$ as the corresponding
$R$-matrix.

\subsection{Statistics of Quantum Fields}

A quantum field on evaluation at a spacetime point (or more generally
on pairing with a test function) gives an operator acting on a Hilbert
space. A field at $x_1$ acting on the vacuum gives a one-particle
state centered at $x_1$. When we write $\Phi(x_1)\,\Phi(x_2)$ we mean
$(\Phi\otimes\Phi)(x_1,x_2)$. Acting on the vacuum we generate a
two-particle state, where one particle is centered at $x_1$ and the
other at $x_2$. (We retain just the creation operator part of $\Phi$
here.) Notice that it just involves evaluation of the two functions
in the tensor product and {\it not} a multiplication map as we get a
function of two variables. On the other hand the star product is a map
from $\mathcal A_\theta \otimes \mathcal A_\theta$ to $\mathcal
A_\theta$ and gives a function of a single variable. Hence there is no
reason a priori to put a star-like operation between
$\Phi(x_1)\,\Phi(x_2)$. We will have more to say about this in Section
8.

If $a_p$ is the annihilation operator of the second-quantized field
$\Phi(x)$, we want, as in standard quantum field theory,
\begin{eqnarray} 
\langle 0 |\Phi^{(-)}(x) a^\dagger_p |0\rangle &=& e_p(x), \\
\frac{1}{2}\langle 0 |\Phi^{(-)}(x_1) \Phi^{(-)}(x_2) a^\dagger_q
a^\dagger_p |0\rangle &=& \left(\frac{{\bf 1} \pm
  \tau_\theta}{2}\right)(e_p \otimes e_q)(x_1,x_2) \nonumber \\
&\equiv& (e_p \otimes_{S_\theta,A_\theta} e_q)(x_1,x_2) 
\label{tbasis} 
\end{eqnarray} 
[We suppress spin indices. Also here we retain only the annihilation
part of the field in $\Phi^{(-)}$]. Note the reversal of $p$ and $q$
from LHS to RHS of (\ref{tbasis}). This is the standard prescription
used to establish the connection between quantum field operators and
(multi-)particle wavefunctions. The correctness of this prescription
can be verified by applying it to the fermionic case, for $\theta^{\mu
\nu}=0$.

This compatibility between twisted statistics and Poincar\'e invariance
has profound consequences for commutation relations. For example when
the states are labeled by momenta, we have, from exchanging $p$ and
$q$ in (\ref{tbasis})
\begin{equation}
|p, q\rangle_{S_\theta,A_\theta} =\ \pm\,e^{ i \theta^{\mu\nu}p_\mu
 q_\nu}\,|q,p \rangle_{S_\theta,A_\theta} 
\end{equation}
This is the origin of the commutation relation
\begin{equation}
a_p^\dagger\,a_q^\dagger\,= \pm e^{ i \theta^{\mu\nu}p_\mu
  q_\nu}\,a_q^\dagger\,a_p^\dagger \, . 
\end{equation}

The adjoint relation 
\begin{equation} 
a_p a_q =  \pm e^{ i \theta^{\mu \nu} p_\mu q_\nu} a_q a_p
\end{equation} 
also follows from the complex conjugate of (\ref{tbasis}) on using
(\ref{tplane}).

The statistics induced on the free quantum fields by (\ref{tbasis}) is
given, on using (\ref{tplane}), by
\begin{equation}
\Phi^{(-)}(x_1)\Phi^{(-)}(x_2)\,=\,\pm
  e^{i\theta^{\mu\nu}\frac{\partial}{\partial x_2^\mu} 
  \frac{\partial}{\partial x_1^\nu}}\,\Phi^{(-)}(x_2) \Phi^{(-)}(x_1) \,.  
\label{spacetwist}
\end{equation}
Any quantization has to be compatible with the above statistics of the
fields.

So far we have had no occasion to use the algebraic properties of
$\mathcal A_\theta$. All we have used is the assumption that the
symmetry of the theory is the twisted Poincar\'e group symmetry. That,
of course, was forced on us from automorphism properties of $\mathcal
A_\theta$.

\section{Choice of Interaction Hamiltonian}

It was claimed by \cite{zahn} that the removal of UV-IR mixing in
noncommutative theories may be due to an inappropriate choice of the
interaction Hamiltonian. Here we point out that our choice of the
Hamiltonian is forced on us from the requirement of twisted Poincar\'e
invariance.

The interaction Hamiltonian is built out of fields. We need a
multiplication map to write down a Hamiltonian density starting from
fields, as it is a scalar function of just one variable. Also in order
to have twisted Poincar\'e invariance, one has to ensure that the
Hamiltonian density transforms like a scalar field.  This will only
happen if we choose a star product (twisted multiplication map)
between the fields to write down the Hamiltonian density. Hence a
generic interaction Hamiltonian density involving just one hermitean
spin zero field (for simplicity) is
\begin{equation}
\mathcal H_I(x)\ =\ \Phi(x)\ast\Phi(x)\ast\cdots\ast\Phi(x)
\end{equation}
where $\Phi(x)$ obeys twisted statistics. This form of Hamiltonian and
the twisted statistics of the fields is all that is needed to show
that there is no UV-IR mixing in this theory \cite{oeckl1,bpq}.

\section{On the Invariance of Correlation Functions}

{\it i) The Twisted Action on the Tensor Product of Plane Waves}:

As a preliminary to the calculations, we first consider the actions of
the twisted coproduct of the Poincar\'e group on the tensor products
of plane waves.

On a single plane wave, the Lorentz transformation $\Lambda$
and translation $P_\mu$ acts according to
\begin{eqnarray} 
(\Lambda e_p)(x) &=& e_p (\Lambda^{-1}x) = e_{\Lambda p}(x), \nonumber
  \label{Lsingle} \\
(P_\mu e_p)(x) &=& -p_\mu e_p(x)
\end{eqnarray} 
where we used $\Lambda^{-1} = \Lambda^T$ and $P_\mu = -i\partial_\mu$. 
Hence
\begin{equation} 
\Lambda e_p = e_{\Lambda p}, \quad \partial_\mu e_p = -i p_\mu e_p \,.
\end{equation} 

Let $U$ denote the representation of the (enveloping algebra of the)
Poincar\'e group on arbitrary tensor products of plane waves. The
latter respond to translations in the usual manner, so we focus on
Lorentz transformations $\Lambda$. On $e_k$, the action of
$U(\Lambda)$ is as in (\ref{Lsingle}):
\begin{equation} 
U(\Lambda)e_k = e_{\Lambda k} \, .
\end{equation} 
On $e_{k_1} \otimes e_{k_2}$, we must find the action using the
coproduct:
\begin{eqnarray} 
U(\Lambda) e_{k_1} \otimes e_{k_2} &=& \Delta_\theta (\Lambda) e_{k_1}
\otimes e_{k_2} \nonumber \\
&=& e^{-\frac{i}{2} \partial_\mu \theta^{\mu \nu} \otimes
  \partial_\nu} (\Lambda \otimes \Lambda) e^{\frac{i}{2} \partial_\mu
  \theta^{\mu \nu} \otimes \partial_\nu} e_{k_1} \otimes e_{k_2}
\nonumber \\
&=& e_{\Lambda k_1} \otimes \underbrace{e^{-\frac{1}{2} (\Lambda k_1)_\mu
  \theta^{\mu \nu} \partial_\nu} \Lambda e^{\frac{1}{2}
  k_{1\mu}\theta^{\mu \nu}\partial_\nu}}_{\Lambda_1} e_{k_2} \\ 
&=& e^{\frac{i}{2} k_1 \cdot \delta_\Lambda \theta \cdot k_2} e_{\Lambda k_1}
\otimes e_{\Lambda k_2}, \nonumber 
\end{eqnarray}
where
\begin{equation}
k_1 \cdot \delta_\Lambda \theta \cdot k_2  \equiv k_{1 \mu} (\delta_\Lambda
\theta)^{\mu \nu} k_{2 \nu}, \quad \delta_\Lambda \theta \equiv \Lambda^{-1}
\theta \Lambda - \theta.
\end{equation}

The action on $e_{k_1} \otimes e_{k_2}\otimes e_{k_3}$ is found using
the coproduct on $\Lambda_1$:
\begin{eqnarray} 
\Delta_\theta ({\Lambda_1}) &=& \left( e^{-\frac{1}{2} {(\Lambda
    k_1)}_\mu \theta^{\mu \nu} 
    \partial_\nu} \otimes e^{-\frac{1}{2}{(\Lambda k_1)}_\mu
    \theta^{\mu \nu} \partial_\nu} \right) \left( e^{-\frac{i}{2}
    \partial_\mu \theta^{\mu \nu}\otimes \partial_\nu} \Lambda \otimes
    \Lambda e^{\frac{i}{2} \partial_\mu \theta^{\mu \nu} \otimes 
      \partial_\nu} \right) \times \nonumber \\
&& \times \left( e^{\frac{1}{2} k_{1\mu}
    \theta^{\mu \nu} \partial_\nu} \otimes e^{\frac{1}{2}k_{1\mu}
    \theta^{\mu \nu} \partial_\nu}\right) \, .
\end{eqnarray} 
It gives 
\begin{equation} 
U(\Lambda) e_{k_1} \otimes e_{k_2} \otimes e_{k_3} = e_{\Lambda k_1}
\otimes \Delta_\theta (\Lambda_1)\left(e_{k_2} \otimes e_{k_3}\right)
\end{equation} 
where
\begin{eqnarray} 
\Delta_\theta (\Lambda_1)\left(e_{k_2} \otimes e_{k_3}\right) &=&
e^{\frac{i}{2} k_1 \cdot \delta_\Lambda \theta \cdot k_2} e_{\Lambda
  k_2} \otimes \Lambda_2 e_{k_3}, \nonumber \\
\Lambda_2 &=& e^{-\frac{1}{2}(\Lambda k_1 + \Lambda k_2)_\mu \theta^{\mu
    \nu} \partial_\nu} \Lambda e^{\frac{1}{2} (k_1 + k_2)_\mu \theta^{\mu
    \nu} \partial_\nu} \, . 
\end{eqnarray} 
Thus
\begin{equation} 
U(\Lambda) e_{k_1} \otimes e_{k_2} \otimes e_{k_3} = e^{\frac{i}{2}k_1
  \cdot \delta_\Lambda \theta \cdot k_2 +\frac{i}{2}(k_1 +
  k_2)\cdot \delta_\Lambda \theta \cdot k_3} e_{\Lambda k_1} \otimes
  e_{\Lambda k_2} \otimes e_{\Lambda k_3}\, .
\end{equation} 

The action on $e_{k_1} \otimes e_{k_2} \otimes e_{k_3} \otimes
e_{k_4}$ is found by splitting $\Lambda_2$ again with a
$\Delta_\theta$. In this way we see that in general,
\begin{equation} 
U(\Lambda)e_{k_1} \otimes e_{k_2} \ldots \otimes e_{k_N} = e^{\frac{i}{2}k_1
  \cdot \delta_\Lambda \theta \cdot k_2 +\frac{i}{2}(k_1 +
  k_2)\cdot \delta_\Lambda \theta \cdot k_3 + \ldots (k_1 + k_2 \ldots +
  k_{N-1})\cdot \delta_\Lambda \cdot \theta k_N} e_{\Lambda k_1}
  \otimes e_{\Lambda k_2} \ldots \otimes e_{\Lambda k_N} \, .
\label{twistedN}
\end{equation} 

\noindent{\it ii) Correlation Functions of NCQFT with Untwisted Statistics:}

Consider the scalar field theory on the GM plane with the Lagrangian
(density)
\begin{equation} 
{\cal L}_\ast = \frac{1}{2} \partial_\mu \Phi \ast \partial^\mu \Phi -
\frac{1}{2} m^2 \Phi \ast \Phi - \frac{\lambda}{4!}  \Phi \ast \Phi
\ast \Phi \ast \Phi \, ,
\label{starLag}
\end{equation} 
where $\Phi^\dagger = \Phi$. Since statistics is not twisted, the
annihilation and creation operators $c_p, c^\dagger_p$ of $\Phi$ are
those for $\theta^{\mu \nu}=0$. 

The correlation functions of (\ref{starLag}) are not Lorentz-invariant
under the twisted coproduct. It is enough to prove this result for the
free field theory where $\lambda =0$. 

The correlation functions for the product of an odd number of fields
is zero. We show now that the four-point function is not
Lorentz-invariant under the twisted coproduct. That can be adapted to
show that the two-point function {\it is} Lorentz
invariant. (Translational invariance is preserved by both untwisted
and twisted statistics.)

The scalar field has the mode expansion
\begin{equation} 
\Phi(x) = \int \frac{d^3p}{(2\pi)^{3/2}(2p_0)} \left( c_p e_p(x) +
c_p^\dagger e_{-p}(x)\right)
\end{equation} 
where $p_0 = +\sqrt{|\vec{p} |^2 + m^2}$ and $c_p$ and $c^\dagger_p$ are
the annihilation-creation operators for $\theta^{\mu \nu}=0$:
\begin{eqnarray} 
{[}c_p, c_k] &=& 0 = [c^\dagger_p,c^\dagger_k], \nonumber \\
{[}c_k , c^\dagger_k] &=& 2p_0 \delta^{3}(p-k)\, .
\end{eqnarray} 
The four point function in this case, with no statistics twist, is
\begin{eqnarray}
\langle 0|\Phi(x_1)\Phi(x_2)\Phi(x_3)\Phi(x_4)|0 \rangle &=&
D(x_1-x_2) \ D(x_3-x_4) + D(x_1-x_3) \ D(x_2-x_4) \nonumber \\
&& + D(x_1-x_4)\ D(x_2-x_3) \nonumber \\
&\equiv& I + II +III \, ,\label{4ptfn}\\
D(x) &=& \int\frac{d^3p}{(2\pi)^3(2p_0)}\,e^{-ip\cdot x}  = D(\Lambda x)
\, .\label{prop}
\end{eqnarray} 
We now show that $I$ and $III$ are invariant (for the twisted coproduct),
but not $II$.

Consider $I$:
\begin{equation} 
I = \frac{1}{(2\pi)^6}\int \left(\prod_i
\frac{d^3p_i}{(2p_{i0})}\right) e_{p_1}(x_1) e_{-p_2}(x_2)
e_{p_3}(x_3) e_{-p_4}(x_4) (2p_{10})(2p_{30}) \delta^{3}(p_1 - p_2)
\delta^{3} (p_3 - p_4)\, .
\label{I}
\end{equation} 
Applying (\ref{twistedN}) with $k_1=p_1,k_2=-p_2,k_3=p_3,k_4=-p_4$, we
find that the phase in (\ref{twistedN}) becomes 1 because of the
$\delta$-functions and that
\begin{equation} 
\Lambda: I \rightarrow D(\Lambda^{-1}(x_1-x_2))
D(\Lambda^{-1}(x_3-x_4)) = I\, . 
\end{equation} 
A similar calculation shows the Lorentz invariance of $III$.

Now consider $II$:
\begin{equation} 
II = \frac{1}{(2\pi)^6}\int \left(\prod_i
\frac{d^3p_i}{(2p_{i0})}\right)e_{p_1}(x_1) e_{p_2}(x_2) e_{-p_3}(x_3)
e_{-p_4}(x_4) (2p_{10})(2p_{20}) \delta^{3}(p_1 - p_3) \delta^{3} (p_2
- p_4).  
\label{II}
\end{equation} 
So with $k_1=p_1,k_2=p_2,k_3=-p_3,k_4=-p_4$ the phase becomes
$e^{\frac{i}{2} p_1 \cdot \delta_\Lambda \theta \cdot p_2}$ and
\begin{equation} 
\Lambda : II \rightarrow \int \frac{d^3p_1 d^3p_2}{(2\pi)^6
  (2p_{10})(2p_{20})} e^{i p_1 \cdot \delta_\Lambda \theta \cdot
  p_2}e^{i (\Lambda p_1)\cdot (x_1-x_3)}e^{i (\Lambda p_2)\cdot
  (x_2-x_4)} \neq II .
\label{IInon}
\end{equation} 
It is not Lorentz-invariant.
\vspace{5mm}

\noindent{\it iii) Correlation Functions of NCQFT with Twisted Statistics:}

In this case the free field is
\begin{equation}
\Phi(x)\,=\,\int \frac{d^3p}{(2\pi)^{3/2}(2p_0)}\,( a_p e_p(x)
+  a_p^\dagger e_{-p}(x) )\, .
\end{equation}
Let ${\cal P}_\mu$ be the Fock space momentum operator:
\begin{equation} 
{\cal P}_\mu = \int \frac{d^3p}{2p_0} p_\mu c^\dagger_p c_p \, .
\end{equation} 
Then, as shown in \cite{bmpv,bmqt}, the operators $a_p,a^\dagger_p$
can be written as follows:
\begin{equation} 
a_p = c_p e^{-\frac{i}{2}p_\mu \theta^{\mu \nu}{\cal P}_\nu}, \quad
a^\dagger_p  = c^\dagger_p e^{+\frac{i}{2}p_\mu \theta^{\mu \nu}{\cal
    P}_\nu} \, . 
\end{equation} 
Using (\ref{I}) and (\ref{II}), we calculate the four-point function
with twisted statistics:
\begin{eqnarray}
&&\langle 0|\Phi(x_1)\Phi(x_2)\Phi(x_3)\Phi(x_4)|0 \rangle = I+III+
\nonumber \\
&&\frac{1}{(2\pi)^6} \int\prod_i \frac{d^3p_i}{(2p_{i0})} e^{i p_{1\mu}
  \theta^{\mu\nu} p_{2\nu}} e_{p_1}(x_1)e_{-p_2}(x_2)e_{p_3}(x_3)
e_{-p_4}(x_4) \times \nonumber \\
&&(2p_{10})(2p_{20})
\delta^{3}(p_1-p_3)\delta^{3}(p_2-p_4). \label{T4ptfn} \\
&\equiv& I + III + II'
\end{eqnarray}
where $I$ and $III$ are Poincar\'e invariant as shown before. As for
$II'$, we find, using (\ref{twistedN}) with
$k_1=p_1,k_2=-p_2,k_3=p_3,k_4=-p_4$ and the $\delta$-functions,
\begin{eqnarray} 
&&\Lambda: II' \rightarrow \frac{1}{(2\pi)^6} \int\prod_i
\frac{d^3p_i}{(2p_{i0})} e_{ \Lambda p_1}(x_1)e_{-\Lambda
  p_2}(x_2)e_{\Lambda p_3}(x_3) e_{-\Lambda p_4}(x_4) \nonumber \\
&& e^{i p_{1\mu}\theta^{\mu \nu}p_{2\nu}} e^{i p_1 \cdot
  \delta_\Lambda \theta \cdot p_2}(2p_{10})(2p_{20})
\delta^{3}(p_1-p_3)\delta^{3}(p_2-p_4). 
\end{eqnarray}  
Since
\begin{equation} 
e^{i p_{1\mu}\theta^{\mu \nu}p_{2\nu}} e^{i p_1 \cdot \delta_\Lambda
  \theta \cdot p_2} = e^{i (\Lambda p_1)_\mu \theta^{\mu\nu}(\Lambda
  p_2)_\nu} 
\end{equation} 
the Poincar\'e invariance of $II'$ also follows. The phase $e^{i
p_{1\mu}\theta^{\mu \nu}p_{2\nu}}$ in (\ref{T4ptfn}) which comes from
twisted statistics is essential to reach this conclusion.

\section{Functional Integral}

We saw above that in order to have twisted Poincar\'e invariance in a
quantum theory, we must also have twisted statistics. This has
implications for a functional integral formulation of the quantum
theory too. This is because statistics of the fields is an input in a
functional integral. For example, in the case of usual fermions,
statistics is not derived from functional integral, but is rather
inferred from other considerations and then built into the functional
integral by use of anticommuting classical fields.

Similarly, in order to construct a functional integral which gives a
twisted Poincar\'e invariant quantum theory, we must use the correct
statistics as an input and construct the functional integral out of
classical fields which obey the twisted statistics. In particular its
full measure consists of tensor products of individual measures at
different points and the individual measures must obey twisted
statistics among themselves in order for the total measure to be
Poincar\'e invariant. This again is in analogy to the case of
fermions, where individual measures anticommute among themselves. We
will not go here into the full details of the construction of the
functional integral which gives the twisted quantum field theory. It
has been done by Oeckl \cite{oeckl2}. It
will suffice here to show that the conventional functional integral does
not give a twisted Poincar\'e invariant theory.

The following functional integral was considered by \cite{tureanu} and
claimed to be twist-Poincar\'e invariant:
\begin{equation}
W\ =\ \int\ {\displaystyle \prod_x} \mathcal D(\phi(x))\,e^{i\int
d^4x\,\mathcal L_\ast (x)}
\end{equation}
where $\mathcal L_\ast$ is for example the star-Lagrangian (density)
\begin{equation}
\mathcal L_\ast
(x)\,=\,\frac{1}{2} \partial_\mu\phi(x)\ast \partial^\mu
\phi(x)\,-\,\frac{1}{2} m^2\phi(x)\ast\phi(x) \,-\,
\frac{\lambda}{4!}\phi(x)\ast\phi(x)\ast \phi(x)\ast\phi(x)
\end{equation}
and $\mathcal D(\phi(x))$ is the usual measure.

With the functional integral defined with this measure, we obtain
conventional quantization of noncommutative field theory with no
statistics twist, and its Feynman rules.

But this measure is not invariant under the twisted Poincar\'e group. We
can show this by a simple argument.

Consider
\begin{eqnarray}
\lefteqn{\int\ {\displaystyle \prod_x} \mathcal
D(\phi(x))\,\phi(x_1)\phi(x_2)\phi(x_3)\phi(x_4)\,e^{i\int
d^4x\,\mathcal L_\ast (x)}}\nonumber\\
&&\qquad\qquad\qquad\phantom m
=\,\langle 0|T\{\Phi(x_1)\Phi(x_2)\Phi(x_3)\Phi(x_4)\}|0 \rangle \, .
\end{eqnarray}
It is enough to consider $\lambda =0$. Let us suppose for convenience
that $x_1^0 \, > \, x_2^0 \,> \,x_3^0 \, > \,x_4^0\,$.  Then
\begin{equation}
\langle 0|T\{\Phi(x_1)\Phi(x_2)\Phi(x_3)\Phi(x_4)\}|0 \rangle\,
=\,\langle 0|\Phi(x_1)\Phi(x_2)\Phi(x_3)\Phi(x_4)|0 \rangle \,.
\end{equation}
which is the same as (\ref{4ptfn}). But we saw above that
\begin{equation}
\langle 0|\Phi(x_1)\Phi(x_2)\Phi(x_3)\Phi(x_4)|0 \rangle\,\neq\,
\Lambda \triangleright
\langle 0|\Phi(x_1)\Phi(x_2)\Phi(x_3)\Phi(x_4)|0 \rangle \, .
\end{equation}
Hence it follows that

\begin{eqnarray}
\lefteqn{\int\ {\displaystyle \prod_x} \mathcal
D(\phi(x))\,\phi(x_1)\phi(x_2)\phi(x_3)\phi(x_4)\,e^{i\int
d^4x\,\mathcal L_\ast (x)}}\nonumber\\
&&\qquad\qquad\qquad\phantom m \neq\,\int\ {\displaystyle \prod_x}
\mathcal D(\phi(x))\, \Lambda \triangleright
(\phi(x_1)\phi(x_2)\phi(x_3)\phi(x_4))\,e^{i\int d^4x\,\mathcal
L_\ast (x)}
\end{eqnarray}
showing that the measure is not twist-Poincar\'e invariant.

\section{Locality}

{\it i) $\theta^{\mu\nu} \neq 0$,Untwisted Statistics:}

The conventional quantization a scalar field on the noncommutative
plane leads to non-local physics. However this non-locality is due to
nonlocal interaction terms and does not show up in the free theory. As
remarked earlier the free theory is identical to the scalar field
theory for $\theta^{\mu \nu}=0$.
\vspace{5mm}

\noindent {\it ii) $\theta^{\mu\nu} \neq 0$, Twisted Statistics:}

The situation is quite different when one quantizes using twisted
statistics. In this case, even the free theory is non-local. We have
\begin{eqnarray}
\lefteqn{[\Phi(x)\,,\,\Phi(y)]=} \nonumber \\
&&\int\frac{d^3pd^3k} {(2\pi)^3 (2p_0)(2 k_0)}\big[\,e^{-i(p\cdot x+k\cdot
y)}(1-e^{-i\theta^{\mu\nu}p_\mu k_\nu})\,a_p a_k
+ e^{i(p\cdot x+k\cdot
y)}(1-e^{-i\theta^{\mu\nu}p_\mu k_\nu})\,a_p^\dagger a_k^\dagger\nonumber\\
&&+e^{-i(p\cdot x-k\cdot y)}\{(1-e^{i\theta^{\mu\nu}p_\mu k_\nu}) a_p
a_k^\dagger\ - (2p_0)\delta^3(p-k)\}\nonumber\\ 
&&+e^{i(p\cdot x-k\cdot y)}\{(1-e^{i\theta^{\mu\nu}p_\mu
k_\nu})\,a_p^\dagger a_k\ + (2p_0)\delta^3(p-k)\}\big]
\end{eqnarray}
This operator is not zero when $x$ and $y$ are space-like separated.
For example, we can calculate it between two single-particle momentum
eigenstates $|q\rangle$ and $|r\rangle$. We have
\begin{eqnarray} 
\langle q|[\Phi(x),\Phi(y)]|r \rangle &=&
(e^{i\theta^{\mu\nu}q_\mu r_\nu}-1) (e^{-i r\cdot
x + i q\cdot y} - e^{i q\cdot x- i r\cdot y}) \nonumber \\
&+& (2q_0) \delta^3(q-r)[D(x-y)-D(y-x)]
\end{eqnarray} 
where $D(x-y)$ was defined in (\ref{prop}). The last two terms
together vanish for space-like separations, but the first term is in
general nonzero for $q\neq r$.

Although the free theory is (twisted) Poincar\'e invariant, it is
non-local. Hence the spin-statistics theorem does not apply to it and
there is no internal inconsistency coming from this theorem.

\section{On Twisted Tensor Product}

In \cite{tureanu}, it has been suggested that the $\ast$-product and
the twist of statistics are one and the same, and that considering
both separately is what led to the result about the absence of UV-IR
mixing.

We feel that this remark is incorrect. It is well-known in Hopf
algebra theory \cite{majid} that the coproduct on a (quasi-triangular)
Hopf algebra is associated with an ``$R$-matrix'' and that the latter fixes
statistics. In our case, $R=\mathcal F^{-2}$ and that gives the
representation of the permutation group via (\ref{newstat}).

Incidentally, a ``twisted'' tensor product has been used in
\cite{tureanu} in connection with the Drinfel'd twist. Its connection
to the $\ast$-product is vague at best. It leads to twisted
statistics, but not the correct one. We can see this as follows.

The ``twisted'' tensor product considered is
\begin{equation} 
\Phi^{(+)}_0 \otimes_T \Phi^{(+)}_0 \equiv e^{\frac{i}{2} \partial_\mu
  \otimes \theta^{\mu \nu} \partial_\nu} \Phi^{(+)}_0 \otimes
  \Phi^{(+)}_0
\end{equation} 
where the field $\Phi^{(+)}_0$ is the creation part (say) of a free
field constructed from the standard creation and annihilation
operators in the usual manner. We have,
\begin{equation} 
\Phi^{(+)}_0(x) \Phi^{(+)}_0(y) = \Phi^{(+)}_0(y) \Phi^{(+)}_0(x)
\end{equation} 
so that 
\begin{eqnarray} 
(\Phi^{(+)}_0 \otimes_T \Phi^{(+)}_0)(x,y) &=& \left(e^{\frac{i}{2}
  \partial_\mu \otimes \theta^{\mu \nu} \partial_\nu}\right)
  (\Phi^{(+)}_0 \otimes \Phi^{(+)}_0)(x,y) \\  
&=&\exp \left(\frac{i}{2} \frac{\partial}{\partial x^\mu} \theta^{\mu
  \nu} \frac{\partial}{\partial y^\nu} \right) \Phi^{(+)}_0(x) 
  \Phi^{(+)}_0(y) \\ 
&=& \exp \left(-\frac{i}{2} \frac{\partial}{\partial y^\mu}
  \theta^{\mu \nu} \frac{\partial}{\partial x^\nu} 
  \right) \Phi^{(+)}_0(y) \Phi^{(+)}_0(x) \\ 
&=& e^{-i \partial_\mu \otimes \theta^{\mu \nu} \partial_\nu}
  (\Phi^{(+)}_0 \otimes_T \Phi^{(+)}_0)(y,x)
\end{eqnarray} 
This does not agree with (\ref{spacetwist}).

{\bf Acknowledgments}: The work of APB, BQ and AP is supported in part
by DOE under grant number DE-FG02-85ER40231.

\end{document}